\documentclass{article}
\usepackage{spconf,amsmath,graphicx}
\usepackage{algorithm}
\usepackage{algpseudocode}
\usepackage{booktabs}
\usepackage{amsfonts}
\usepackage{hyperref}

\title{CatNet: music source separation system with mix-audio augmentation}
\name{Xuchen Song*, Qiuqiang Kong*\thanks{* Equal contribution.}, Xingjian Du, Yuxuan Wang}
\address{ByteDance, Shanghai, China \\ \{xuchen.song, kongqiuqiang, duxingjian.real, wangyuxuan.11\}@bytedance.com}

\begin{document}
%
\maketitle

\begin{abstract}
Music source separation (MSS) is the task of separating a music piece into individual sources, such as vocals and accompaniment. Recently, neural network based methods have been applied to address the MSS problem, and can be categorized into spectrogram and time-domain based methods. However, there is a lack of research of using complementary information of spectrogram and time-domain inputs for MSS. In this article, we propose a CatNet framework that concatenates a UNet separation branch using spectrogram as input and a WavUNet separation branch using time-domain waveform as input for MSS. We propose an end-to-end and fully differentiable system that incorporate spectrogram calculation into CatNet. In addition, we propose a novel mix-audio data augmentation method that randomly mix audio segments from the same source as augmented audio segments for training. Our proposed CatNet MSS system achieves a state-of-the-art vocals separation source distortion ratio (SDR) of 7.54 dB, outperforming MMDenseNet of 6.57 dB evaluated on the MUSDB18 dataset.

\end{abstract}
%
%
%
%

%
\section{Introduction}\label{sec:headings}
Music source separation (MSS) is the task of separating a music piece into individual sources, such as vocals and accompaniment. MSS has several applications in music information retrieval \cite{casey2008content}, including melody extraction, vocal pitch correction and music transcription. Single-channel MSS is a challenging problem because the number of sources to be separated is larger than the number of channels of mixtures. For example, audio recordings from the MUSDB18 dataset \cite{musdb18} have two channels, while the separation task is to separate audio recordings into four channels including vocals, drums, bass and other sources.

Early works of MSS include using unsupervised separation methods, such as harmonic structures to separate sources \cite{duan2008unsupervised}, and using non-negative matrix factorizations (NMFs) to factorize multi-channel or single-channel spectrograms \cite{ozerov2009multichannel, weninger2014discriminative}. Independent component analysis (ICA) has been proposed for signal separation in \cite{davies2007source}. Recently, neural network based methods have been applied to MSS, such as fully connected neural networks \cite{grais2014deep,rafii2018overview}, recurrent neural networks (RNNs) \cite{huang2015joint, uhlich2017improving, stoter2019open}, and convolutional neural networks (CNNs) \cite{jansson2017singing, hennequin2019spleeter}. Those neural network based methods usually use spectrograms of audio recordings as input features, and build neural networks to separate the spectrograms of target sources. In addition to the spectrogram based methods, several time-domain MSS systems have been proposed, such as WavUNet \cite{stoller2018wave}, Conv-TasNet \cite{luo2019conv} and Demucs \cite{defossez2019demucs}. Those time-domain MSS systems use the waveform of audio recordings as input features, and build neural networks to separate the waveforms of target sources. However, many of those time-domain MSS systems do not outperform spectrogram based MSS systems.

In this article, we propose an end-to-end MSS system called CatNet to combine the advantages of spectrogram and time-domain systems for MSS, where \textit{CatNet} is the abbreviation of \textit{concatenation of networks}. A CatNet consists of a UNet \cite{jansson2017singing} branch using spectrogram as input and a WavUNet \cite{stoller2018wave} branch using time-domain waveform as input. To build a consistent input format for the UNet branch and the WavUNet branch, we wrap Fourier and inverse Fourier transform into differentiable matrix multiplication operations. Then, we sum the UNet and WavUNet outputs in the time-domain to obtain separated sources. Therefore, CatNet utilizes the complementary information of spectrogram and time-domain systems for MSS, and is fully differentiable which can be trained in an end-to-end way.

Data augmentation is important to improve the performance of MSS systems when the duration of training data is limited to hours such as the MUSDB18 dataset \cite{musdb18}. However, previous data augmentation methods \cite{uhlich2017improving, pretet2019singing} include adding filters, remixing audio recordings, swapping left and right channels, shifting pitches, scaling and stretching audio recordings have marginal improvement to the performance of MSS systems \cite{uhlich2017improving, pretet2019singing}. We propose a novel mix-audio data augmentation method that randomly mix audio segments from the same source as augmented audio segments.

This paper is organized as follows, Section 2 introduces neural network based MSS systems. Section 3 introduces our proposed CatNet system. Section 4 introduces our proposed mix-audio data augmentation strategy. Section 5 shows experimental results, and Section 6 concludes this work.

\section{Source separation methods}
Previous neural network based MSS systems can be categorized into spectrogram and time-domain based methods. We introduce those systems as follows.

\subsection{Spectrogram Based Methods}\label{section:sp_methods}
Recently, neural networks have been applied to MSS, including fully connected neural networks \cite{grais2014deep, nugraha2016multichannel}, recurrent neural networks (RNNs) \cite{huang2015joint, uhlich2017improving} and convolutional neural networks (CNNs) \cite{jansson2017singing, hennequin2019spleeter, liu2020voice}. A neural network based MSS system learns a mapping from the spectrogram of a mixture to the spectrogram of a target source. To begin with, we denote the mixture of an input as $ x $, and the source to separate as $ s $. We apply a short time Fourier transform (STFT) $ \mathcal{F} $ on $ x $ and $ s $ to calculate their STFT $ X = \mathcal{F}(x) $ and $ S = \mathcal{F}(s) $ respectively. Both $ X $ and $ S $ are complex matrices with shapes of $ T \times F $, where $ T $ and $ F $ are the number of frames and frequency bins respectively. We call the magnitude of $ |X| $ spectrogram. A MSS system takes a spectrogram $ |X| $ as input, and outputs a separated spectrogram $ |\hat{S}| $:
\begin{equation} \label{eq:sp_regression}
|\hat{S}| = f_{\text{sp}}(|X|),
\end{equation}
\noindent where $ f_{\text{sp}} $ is a mapping modeled by a set of learnable parameters. For example, $ f_{\text{sp}} $ can be a RNN, a CNN, or a UNet \cite{jansson2017singing}. In this article, we model (\ref{eq:sp_regression}) with a mask based approach \cite{wang2018supervised}:
\begin{equation} \label{eq:mask}
|\hat{S}| = m \odot |X|,
\end{equation}
\noindent where $ m \in [0, 1]^{T \times F} $ is a mask to be estimated by the RNN, CNN or UNet, and $ \odot $ denotes element-wise multiplication. Then, the separated STFT $ \hat{S} $ can be obtained by:

\begin{equation} \label{eq:sp_istft}
\hat{S} = |\hat{S}| e^{\angle X},
\end{equation}

\noindent where $ \angle X $ is the phase of $ X $. Finally, an inverse STFT (ISTFT) $ \mathcal{F}^{-1} $ is applied to $ \hat{S} $ to output the separated waveform $ \hat{s} $. 

\subsection{Time-domain Based Methods}\label{section:time_methods}
There are several disadvantages of spectrogram based MSS methods. First, spectrogram based MSS systems only predict the spectrogram $ |\hat{S}| $ of sources, but do not estimate the phases of separated sources. Therefore, the performance of those MSS systems are limited. Recently, several works have investigated estimating the phases of sources for MSS \cite{takahashi2018phasenet}. Secondly, spectrogram may not be an optimal representation for a time-domain signal. For example, STFT performs well in analysing stationary signals in a short time window, while signals are not always stationary. Time-domain MSS systems have been proposed to address those problems. Both the input and output of time-domain systems are waveforms instead of spectrograms. Those time-domain systems have the advantage of not needing to estimate phases for MSS, such as WavUNet \cite{stoller2018wave}, Conv-TasNet \cite{luo2019conv}, and Demucs \cite{defossez2019demucs}. Similar to the spectrogram based MSS systems, we build a mapping $ f_{\text{wav}} $ from the mixture signal $ x $ to separate source $ \hat{s} $:
\begin{equation} \label{eq:wave_regression}
\hat{s} = f_{\text{wav}}(x),
\end{equation}
\noindent where $ f_{\text{wav}} $ is a regression function modeled by a set of learnable parameters. Equation (\ref{eq:wave_regression}) does not need to apply STFT and ISTFT for separating sources. In this article, we model $ f_{\text{wav}} $ with a WavUNet \cite{stoller2018wave}.

\section{CatNet for music source separation}\label{section:proposed}

\subsection{CatNet}
The motivation of proposing CatNet is to combine the advantages of spectrogram and time-domain based methods. UNet architectures have shown to perform good in capturing the time-frequency patterns of spectrograms for MSS. However, spectrogram may not be the optimal representation for audio recordings. On the other hand, WavUNet has the advantage of learning adaptive kernels to represent audio recordings in the time-domain waveform. Our proposed CatNet is designed to combine the advantages of UNet and WavUNet by concatenating their time-domain outputs. CatNet consists of two branches. One branch is a UNet that uses differentiable operations to calculate spectrogram, and outputs time-domain waveforms described in Section \ref{section:ete_sp}. The UNet branch is designed to learn robust frequency patterns of sounds. Another branch is a WavUNet that takes the waveform of a mixture and outputs the waveform of separated sources. WavUNet provides flexible and learnable transforms by using one-dimensional convolutions. The UNet branch and WavUNet branch provide complementary information for MSS. We denote the outputs of UNet and WavUNet branches as $ \hat{s}_{\text{U}} $ and $ \hat{s}_{\text{WU}} $ respectively. The output of CatNet can be written as:
\begin{equation} \label{eq:unet_wavunet}
\hat{s} = \hat{s}_{\text{U}} + \hat{s}_{\text{WU}}.
\end{equation}

\noindent Equation (\ref{eq:unet_wavunet}) shows that the output of a CatNet is the addition of the outputs of UNet and WavUNet in the time-domain. In training, the parameters of UNet and WavUNet are optimized jointly. Previous UNet system \cite{jansson2017singing} uses spectrogram as input and output, which can not be used in our end-to-end CatNet system. Therefore, we propose to incorporate UNet into our end-to-end time-domain MSS system as follows.

\subsection{Incorporating Spectrogram into CatNet}\label{section:ete_sp}
STFT can be calculated by applying discrete Fourier transform (DFT) on the frames of audio clips, where each frame contains several audio samples in a short time window. An audio recording $ x $ is split into $ T $ frames, where each frame contains $ N $ samples. We denote the $ t $-th frame as $ x_{t} $, and its DFT as $ X_{t} $. The calculation of DFT can be written as:
\begin{equation} \label{eq:dft}
X_{t} = Dx_{t}, 
\end{equation}
\noindent where $ D $ is an $ N \times N $ complex matrix with elements of $ D_{kn} = e^{-\frac{i2\pi}{N}kn} $. We decompose the DFT matrix $ D $ into a real part $ D_{\text{R}} $ and an imaginary part $ D_{\text{I}} $. Then, equation (\ref{eq:dft}) can be written as:
\begin{equation} \label{eq:dft_ri}
X_{t} = D_{\text{R}}x_{t} + iD_{\text{I}}x_{t}.
\end{equation}
Equation (\ref{eq:dft_ri}) shows that STFT can be calculated by matrix multiplication. We apply one-dimensional convolutional operations to $ x $ to parallel the calculation of STFT over several frames: $ X_{\text{R}} = \text{conv}_{\text{R}}(x) $ and $ X_{\text{I}} = \text{conv}_{\text{I}}(x) $, where $ \text{conv}_{\text{R}} $ and $ \text{conv}_{\text{I}} $ are one-dimensional convolutions with parameters of $ D_{R} $ and $ D_{I} $ respectively. The strides of $ \text{conv}_{\text{R}} $ and $ \text{conv}_{\text{I}} $ are set to the hop samples between adjacent frames. Then, the STFT of $ x $ is calculated by $ X = X_{R} + iX_{I} $. In separation, we calculate the separated spectrogram $ |\hat{S}| $ using the system described in Section \ref{section:sp_methods}. Then, the estimated STFT $ \hat{S} $ can be obtained by:
\begin{equation} \label{eq:phases}
\hat{S} = |\hat{S}|\text{cos}\angle X + i|\hat{S}|\text{sin}\angle X,
\end{equation}
\noindent where $ \angle X $ is the angle of $ X $. Similar to the calculation of STFT, ISTFT can be calculated by applying inverse DFT (IDFT) on the frames of $ \hat{S} $ \cite{wang2015deep}. We denote the $ t $-th frame of $ \hat{S} $ as $ \hat{S}_{t} $, and the $t$-th frame of time-domain IDFT as $ \hat{s}_{t} $. The calculation of $ \hat{s}_{t} $ can be written as:
\begin{equation} \label{eq:idft}
\hat{s}_{t} = D^{-1}\hat{S}_{t},
\end{equation}
\noindent where $ D^{-1} $ is an $ N \times N $ complex IDFT matrix with elements of $ D_{kn} = \frac{1}{N} e^{\frac{i2\pi}{N}kn} $, and $ N $ is the number of samples in a frame. We decompose the IDFT matrix $ D^{-1} $ and estimated $ \hat{S} $ into real and imaginary parts: $ D^{-1} = D^{-1}_{\text{R}} + iD^{-1}_{\text{I}} $ and $ \hat{S}_{t} = \hat{S}_{\text{R},t} + i\hat{S}_{\text{I},t} $. Considering the reconstructed signal is a real signal, the signal $ \hat{s}_{t} $ can be reconstructed by:

\begin{equation} \label{eq:idft_ri}
\hat{s}_{t} = D^{-1}_{\text{R}}\hat{S}_{R,t} - D^{-1}_{\text{I}}\hat{S}_{I,t}.
\end{equation}
\noindent Similar to the calculation of STFT, we apply one-dimensional convolutional operations to $ \hat{S} $ to parallel the reconstruction of time-domain signals $ \hat{s}_{1,...,T} = \{\hat{s}_{1}, ..., \hat{s}_{T}\} $:
\begin{equation} \label{eq:istft}
\begin{matrix}
\hat{s}_{1,...,T} = \text{conv}^{-1}_{\text{R}}(\hat{S}_{R}) - \text{conv}^{-1}_{\text{I}}(\hat{S}_{I}),\\ 
\end{matrix}
\end{equation}
\noindent where $ \text{conv}^{-1}_{\text{R}} $ and $ \text{conv}^{-1}_{\text{I}} $ are one-dimensional convolutions with parameters of $ D^{-1}_{\text{R}} $ and $ D^{-1}_{\text{I}} $ respectively. Finally, a transposed convolution with a stride equivalent to hop samples between adjacent frames is used to reconstruct the time-domain waveform with overlap-add:
\begin{equation} \label{eq:tconv}
\hat{s} = \text{tconv}(\hat{s}_{1,...,T}).
\end{equation}
All STFT, ISTFT and overlap-add operations are built into the end-to-end CatNet, and are fully differentiable.

\section{Mix-audio data augmentation}
The amount of training data is important for training MSS systems. The duration of public available MSS datasets such as MUSDB18 \cite{stoter20182018, musdb18} is limited to a few hours. Data augmentation is a technique to increase the variety of training data. Previous works have applied several data augmentation methods for MSS \cite{uhlich2017improving, pretet2019singing} including adding filters to songs, remixing sources from different songs, swapping left and right channels, shifting pitches, scaling and stretching audio recordings with random amplitudes \cite{uhlich2017improving, pretet2019singing}. However, those data augmentation methods have marginal influence on training MSS systems. In this work, we propose a novel mix-audio data augmentation technique for MSS. Different from previous remixing instruments method \cite{uhlich2017improving, pretet2019singing} that remix instruments from different songs, our proposed mix-audio data augmentation method randomly mixes two audio segments from a \textit{same} source as an augmented segment for that source:
\begin{equation} \label{eq:mix_s}
s_{\text{mix}} = \sum_{j=1}^{J}s_{j},
\end{equation}
\noindent where $ s_{j}, j=1,...,J $ are audio segments from the same source, and $ J $ is the number of sources to be mixed. We denote the mixed source as $ s_{\text{mix}} $. If $ s_{j} $ are vocals, then, their addition $ s_{\text{mix}} $ is also vocals. The mix-audio data augmentation provides a large amount combinations of one source. We denote $ s_{\text{mix}}^{(i)} $ as the $ i $-th source, and $ I $ as the number of sources to separate. The input mixture to a CatNet is calculated by:
\begin{equation} \label{eq:mix_x}
x_{\text{mix}} = \sum_{i=1}^{I}s_{\text{mix}}^{(i)},
\end{equation}
\noindent To explain, the MSS turns to a problem of constructing a mapping $ f $ from $ x_{\text{mix}} $ to separate $ \hat{s}_{\text{mix}} $ by using the CatNet described in (\ref{eq:unet_wavunet}).
With mix-audio data augmentation, separating $ \hat{s}_{\text{mix}} $ from $ x_{\text{mix}} $ is a more challenging problem than that without mix-audio data augmentation. The effectiveness of mix-audio data augmentation comes from that the addition of several audio segments from a same source also belongs to that source. Intuitively, a system that is able to separate multiple vocals from a mixture is also able to separate a single vocal from the mixture.

\section{Experiments}
We experiment our proposed CatNet system on the MUSDB18 dataset \cite{musdb18, stoter20182018}. The MUSDB18 dataset consists of 150 full length music tracks of different genres including isolated vocals, drums, bass and other sources. The training and testing set contains 100 and 50 songs respectively. All music recordings are stereophonic with a sampling rate of 44.1 kHz. In training, music pieces are split into 3-second audio segments. Random track mixing is used in our systems as a default setting. For mix-audio data augmentation, we randomly select audio segments from a source to obtain $ s_{\text{mix}} $ as described in (\ref{eq:mix_s}). Then, we randomly mix $ s_{\text{mix}} $ from different sources to constitute the mixture $ x_{\text{mix}} $ as input described in (\ref{eq:mix_x}).

A CatNet consists of a UNet branch and a WavUNet branch. For the UNet branch, spectrogram is extracted using one-dimensional convolutions described in Section \ref{section:ete_sp} with a Hann window size 2048 and a hop size 441. The UNet branch consists of six encoding and six decoding blocks. Each encoding block consists of two convolutional layers with kernel sizes of $ 3 \times 3 $ and an $ 2 \times 2 $ average pooling layer. Each convolutional layer consists of a linear convolution, a batch normalization \cite{ioffe2015batch} and a ReLU \cite{nair2010rectified} nonlinearity. Each decoder layer consists of two convolutional layers that are symmetric to the encoder. In UNet, the output of each encoder is concatenated with the output of the encoder from the same hierarchy, and is input to a transposed convolutional layer with a kernel size $ 3 \times 3 $ and a stride $ 2 \times 2 $ to upsample the feature maps. The numbers of channels of encoder blocks are 32, 64, 128, 256, 512 and 1024, and the number of channels of decoder blocks are symmetric to encoder blocks. The WavUNet branch consists of six encoder blocks and six decoder blocks. The kernel size of all one-dimensional convolutional layers are 3. The average pooling layers have sizes of 4, and the transpose convolutional layers have strides of 4. The number of channels in the encoder blocks are 32, 64, 128, 256, 512 and 1024, and decoder blocks are symmetric to encoder blocks. We use an Adam optimizer \cite{kingma2014adam} with a learning rate of 0.001, and a batch size of 12 for training. We apply mean absolute error loss function $ \left \| s - \hat{s} \right \|_{1} $ between estimated and target time-domain waveforms for training.

\begin{table}[t]
\centering
\caption{SDRs of previous MSS systems}
\label{table:compare_result}
\begin{tabular}{*{5}{c}}
 \toprule
 & Vocals & Drums & Bass & Other \\
 \midrule
WavUNet \cite{stoller2018wave} & 3.05 & 4.16 & 3.17 & 2.24 \\
Demucs \cite{defossez2019demucs} & 6.21 & 6.50 & 6.21 & 3.80 \\ 
MMDenseNet \cite{takahashi2017multi} & 6.57 & 6.40 & 5.14 & 4.13 \\
DAE \cite{liu2018denoising} & 5.74 & 4.66 & 3.67 & 3.40 \\
 \midrule
 CatNet + aug & 7.54 & 5.85 & 5.01 & 4.67 \\
  \bottomrule
 \end{tabular}
\end{table}

\begin{table}
  \caption{SDRs of our proposed MSS systems with mix-audio data augmentation.}
  \vspace{6pt}
  \label{tab:full_result}
  \centering
  \resizebox{\columnwidth}{!}{%
  \begin{tabular}{lccccc}
    \toprule
    & Vocals & Acc. & Drums & Bass & Other \\
    \midrule
 UNet (sp) & 7.04 & 15.06 & 5.72 & 4.28 & 4.38 \\
 \midrule
 WaveUNet & 5.57 & 12.68 & 5.03 & 5.42 & 3.38 \\
 WaveUNet + aug & 6.08 & 13.43 & 5.15 & 4.65 & 3.49 \\
 UNet (wav) & 7.13 & 15.47 & 5.96 & 4.37 & 4.50 \\
 UNet (wav) + aug & 7.17 & 15.20 & 5.86 & 3.79 & 4.18 \\
 CatNet & 6.96 & 15.13 & 5.91 & 6.10 & 4.99 \\
 CatNet + aug & 7.54 & 15.18 & 5.85 & 5.01 & 4.67 \\
\bottomrule
\end{tabular}}
\end{table}

\subsection{Results}
We evaluate the MSS performance with source to distortion ratio (SDR) \cite{vincent2006performance} using the BSSEval V4 toolkit \cite{stoter20182018}. Median SDR is used for evaluating different MSS systems \cite{stoter20182018}. Table \ref{table:compare_result} shows the results of previous MSS systems. The time-domain system WavUNet \cite{stoller2018wave} achieves a SDR of 3.05 dB in vocals separation, and is improved by Demucs \cite{defossez2019demucs} with a median SDR of 6.21 dB. The spectrogram based method MMDenseNet \cite{takahashi2017multi} outperforms the time-domain methods, with a median SDR of 6.57 dB. A denoising autoencoder (DAE) system \cite{liu2018denoising} achieves a SDR of 5.74 dB. The bottom row shows that our proposed CatNet with mix-audio augmentation achieves a state-of-the-art vocals separation SDR of 7.54, outperforming other systems.

All systems in Table \ref{tab:full_result} are our re-implemented systems. The first row of Table \ref{tab:full_result} shows that UNet system \cite{jansson2017singing} trained with spectrogram loss achieves a vocals SDR of 7.04 dB and an accompaniment SDR of 15.06 dB. The second row shows that WavUNet \cite{stoller2018wave} achieves a vocals separation SDR of 5.57 dB, underperforming the spectrogram based method. The text ``+ aug'' indicates with mix-audio data augmentation. The third row shows that with mix-audio data augmentation, the vocals SDR is improved to 6.08 dB. The fourth row shows that our proposed spectrogram based UNet trained with time-domain MAE loss improves the vocals separation SDR to 7.13 dB, and improves the SDR to 7.17 dB with mix-audio data augmentation. The seventh row shows that our proposed CatNet with mix-audio data augmentation achieves a state-of-the-art vocals separation SDR of 7.54 dB, largely outperforming other systems, indicating the effectiveness of vocals separation with mix-audio data augmentation. Our system also achieves accompaniment, drums, bass and other separation SDRs of 15.18 dB, 5.85 dB, 5.01 dB and 4.67 dB respectively.

\section{Conclusion}
We propose an end-to-end CatNet for music source separation. The CatNet consists of a UNet branch and a WavUNet branch to combine the advantages of both spectrogram and time-domain MSS systems. CatNet is fully differentiable and can be trained in an end-to-end say. We propose a novel mix-audio data augmentation method that randomly mix audio segments from the same source as augmented audio segment. Our proposed CatNet with mix-audio data augmentation system achieves a state-of-the-art vocals separation SDR of 7.54 dB, and an accompaniment separation SDR of 15.18 dB on the MUSDB18 dataset. In future, we will investigate using CatNets for general source separation.

\small
\bibliography{refs}
\bibliographystyle{IEEEbib}

\end{document}